\documentclass[10pt,journal,twocolumn]{IEEEtran}
\usepackage{graphicx}
\usepackage{amsmath}
\usepackage{amssymb}
\usepackage{enumerate}
\usepackage{cleveref}
\usepackage{verbatim}
\usepackage{xcolor}
\usepackage{todonotes}
\usepackage{array}
\usepackage{cancel}
\usepackage{subfiles}

\newcommand{\vdelta}{\mathbf{\Delta}}
\newcommand{\vx}{\mathbf{x}}
\newcommand{\rmd}{\mathrm{d}}

\begin{document}
\title{Image registration and super resolution from first principles}
\author{Colin~B.~Clement,~Matthew~Bierbaum,~and~James~P.~Sethna\\
Laboratory of Atomic and Solid State Physics, Cornell University,\\
Ithaca, New York 14853-2501, USA}
\date{\today}
\maketitle

\begin{abstract} 
    Image registration is the inference of transformations relating noisy and
    distorted images. It is fundamental in computer vision, experimental
    physics, and medical imaging. Many algorithms and analyses exist for
    inferring shift, rotation, and nonlinear transformations between image
    coordinates. Even in the simplest case of translation, however, all known
    algorithms are biased and none have achieved the precision limit of the
    Cramer Rao bound (CRB). Following Bayesian inference, we prove that the
    standard method of shifting one image to match another cannot reach the CRB.
    We show that the bias can be cured and the CRB reached if, instead, we use
    Super Registration: learning an optimal model for the underlying image and
    shifting that to match the data. Our theory shows that coarse-graining
    oversampled images can improve registration precision of the standard
    method. For oversampled data, our method does not yield striking
    improvements as measured by eye. In these cases, however, we show our new
    registration method can lead to dramatic improvements in extractable
    information, for example, inferring $10\times$ more precise particle
    positions.
\end{abstract}


\maketitle
\section{Introduction}

Image registration is the problem of inferring the coordinate transformation
between two (or more) noisy and shifted (or distorted) signals or images.  This
deceptively simple process is fundamental for stereo
vision~\cite{lucas1981iterative}, autonomous vehicles~\cite{wolcott2014visual},
gravitational astronomy~\cite{nicholson1998bayesian}, remote
sensing~\cite{inglada2007analysis,debella2011sub}, medical
imaging~\cite{zollei2003unified,leventon1998multi},
microscopy~\cite{savitzky2018image}, and nondestructive strain
measurement~\cite{kammers2013digital}. At the cutting edge of microscopy, imaging sensitive biological
materials~\cite{bartesaghi20152,bartesaghi2014structure} and metal organic
frameworks~\cite{zhang2018atomic,zhu2017unravelling} with Transmission Electron
Microscopy, requires combining multiple low-dose high-noise images, to obtain a
viable signal without destroying the sample. While most techniques for
registering and combining images are accurate
for low noise, errors significantly larger than theoretical bounds
can occur for a signal-to-noise ratio as low as 20 (noise 5\% of the signal
amplitude); so far a general explanation of this error has been elusive.

Much has been written about the uncertainty of shift estimations by analyzing
the information theoretic limit known as the Cramer-Rao bound
(CRB)~\cite{robinson2004fundamental,yetik2006performance,pham2005performance}.
These works observed that no known estimators achieve the CRB for image
registration. This sub-optimal performance has been blamed on biased
estimators: some claim interpolation errors explain the
bias~\cite{rohde2009interpolation,schreier2000systematic,inglada2007analysis,bailey2005bias}
and others claim that the problem is inherently
biased~\cite{robinson2004fundamental}. More works have explored non-perturbative
estimations of the uncertainty, which yield larger estimates more consistent
with measured error, but also rely on assumptions about the latent
image~\cite{uss2014precise,ziv1969some,xu2009ziv}.

Here we solve these problems by studying the na\"ive maximum likelihood
formulation of image registration. We explore a new derivation of the standard
method (comparing one image to match the other) by integrating out the
underlying true image. We treat the standard method as a statistical field
theory in which two images fluctuate around each other, showing that the shift
uncertainty should scale quadratically with image noise
($\sigma_\Delta\propto\sigma^2$), while the na\"ive CRB is linear
($\sigma_\Delta\propto\sigma$). We also show that bias in image registration is
due to the image edges. Our theory makes the novel prediction that
coarse-graining images can dramatically improve shift precision, which we
confirm numerically. While coarse-graining helps, it requires oversampled images
and knowledge of the highest frequencies of the underlying image. We overcome
this limitation, and reach the true CRB, by shifting a learned model for the
underlying image to match the data.  We use Bayesian model selection to find the
model most supported by the data, effectively learning the amount of necessary
coarse-graining.  We demonstrate the optimality of our new method---called Super
Registration (SR)---with periodic images. We also demonstrate clear improvements
in error and removal of bias for general non-periodic images with Chebyshev
image models.  Finally, we show that particle tracking is 10-20$\times$ more
precise when performed on images combined with SR. We conclude by discussing the
implications of our theory on more general nonlinear registration, and
registration of images captured with different imaging modes.

\begin{figure}
    \includegraphics[width=0.45\columnwidth]{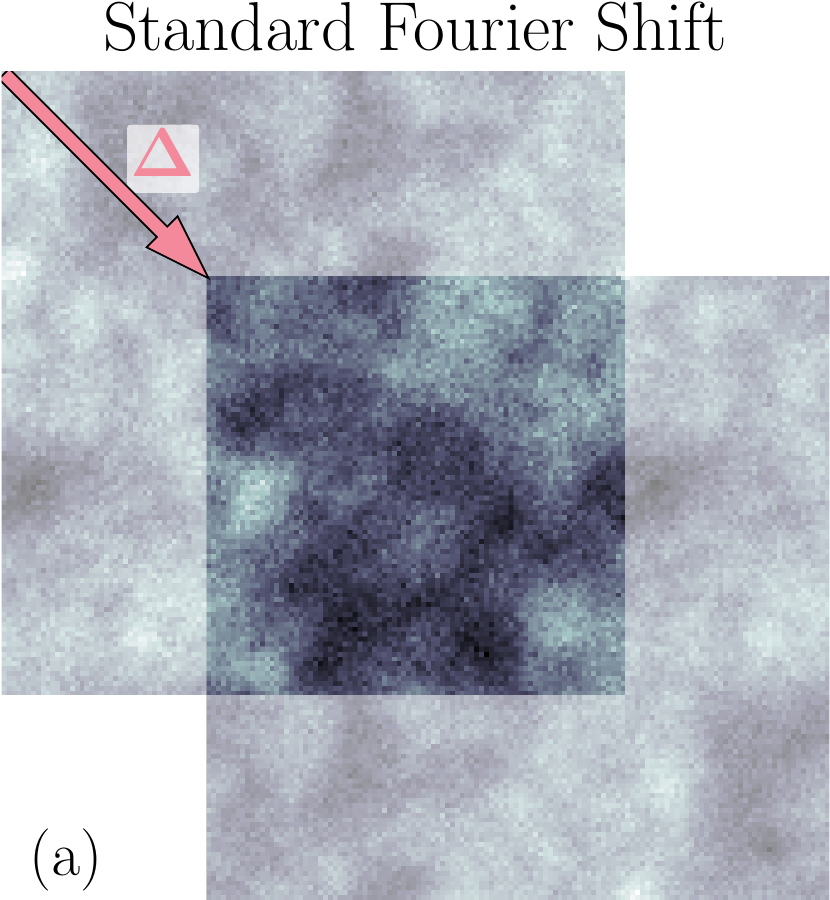}~~~
    \includegraphics[width=0.45\columnwidth]{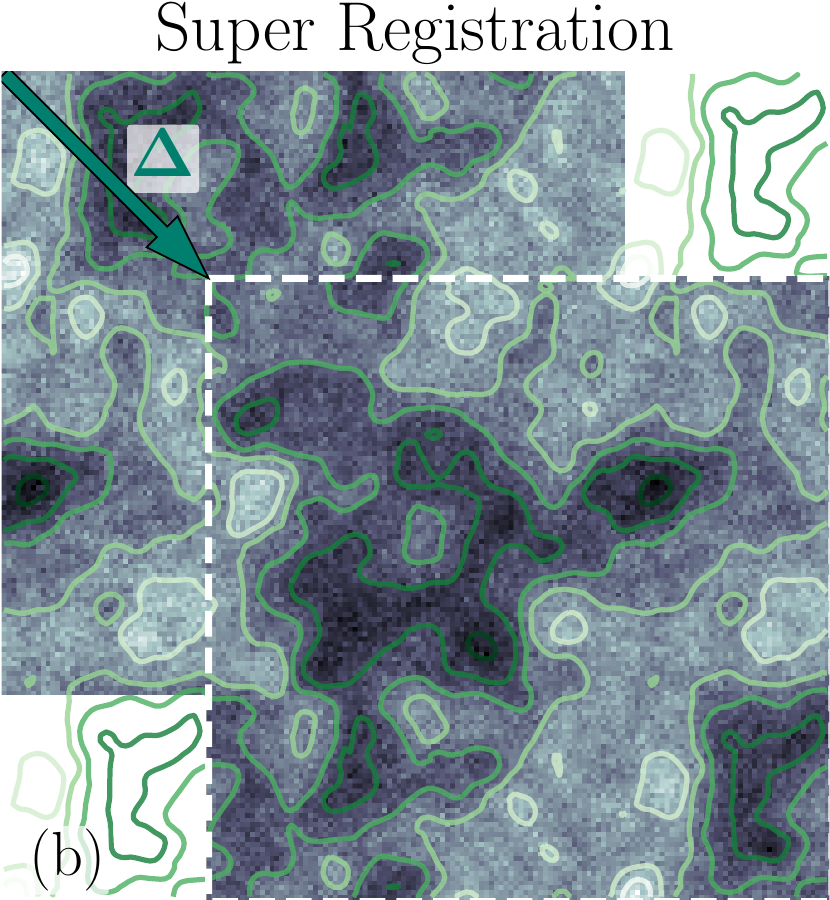}

    \caption{Illustrations of image registration techniques.  (a) A schematic
    of the standard method of image registration which measures the shift
    $\Delta$ between noisy data (grayscale images) by shifting one to match the
    other.  (b) A schematic of our proposed method, Super Registration, which
    infers the shift $\Delta$ instead by learning the underlying image $I$
    (green contours), and shifting the coordinates until the model image best
    fits the data (grayscale images).}

    \label{fig:schematics}
\end{figure}

\section{Theory of image formation}

In this work, image registration will be restricted to the task of inferring a
rigid shift relating two (or more) discretely sampled noisy images with
sub-pixel precision. More general transformations are accommodated by our
subsequent arguments through application of the chain rule. Defining some true
image (latent, to be discovered) intensity function $I(\vx)$ with $\vx\in
\mathbb{R}^2$, we measure at least two images by sampling discretely:
\begin{align}
    \phi_i &= I(\vx_i) + \xi_i\nonumber\\
    \psi_i &= I(\vx_i+\vdelta) + \xi_i,
    \label{eq:model}
\end{align}
where $\phi_i$ is the $i^\mathrm{th}$ pixel of image $\phi$ and $\xi_i$ are white
noise distributed with zero mean and variance $\sigma^2$, and $\vdelta$ is the
shift between the images which we intend to infer.

Equation~\ref{eq:model} is our model, which we can express as
the likelihood $p(\phi,\psi|\vdelta, I)$
of measuring $\phi$ and $\psi$ given $\Delta$ and $I$:
\begin{equation}
    p(\phi,\psi|\vdelta, I) \propto \exp \left(-\frac{1}{2\sigma^2}\left(
       ||\phi - I||^2 + ||\psi-T_\vdelta I||^2 \right)\right),
    \label{eq:likelihood}
\end{equation}
where $||x||^2=\sum_i x_i^2$. $T_\vdelta$ represents the operator which
translates its argument by $\vdelta$, for a continuous image $T_\vdelta I(\vx) =
I(\vx-\vdelta)$. We interpret this distribution as our image model fluctuating
around data. Note that Eq.~\ref{eq:likelihood} accommodates multi-image
registration by multiplying more products of terms comparing images to the
shifted latent image $I$.

In order to infer $\vdelta$ after measuring the images $\phi$ and $\psi$ we must
reverse the conditional probability in Eqn.~\ref{eq:likelihood} using Bayes'
theorem. The posterior (post-measurement) probability 
$p(\vdelta, I|\phi,\psi)$ of $\vdelta$ and $I$ is
\begin{equation}
    p(\vdelta, I|\phi, \psi) = \frac{p(\phi, \psi|\vdelta, I)p(\vdelta, I)}
    {p(\phi, \psi)}.
    \label{eq:posterior}
\end{equation}
$p(\vdelta,I)$ is called the prior probability and $p(\phi,\psi)$ is called the
evidence because, as we later show, it can be interpreted as the probability of
our data given our choice of model.  The task of inferring $\vdelta$ is achieved
by maximizing this posterior probability.  We define the maximum likelihood
estimator of $\vdelta$ to be 
\begin{align}
    \vdelta^\star &= \text{max}_{\vdelta,I}~ p(\vdelta,I|\phi,\psi),\nonumber\\
                  &= \text{max}_{\vdelta,I}~ p(\phi, \psi|\vdelta, I)p(\vdelta, I),
    \label{eq:ml}
\end{align}
where the second line is possible because the evidence is independent of 
$\vdelta$ and $I$.

How accurately should we be able to measure $\vdelta$?  If we assume we know the
underlying image $I$, the answer is given by the Cramer-Rao bound
(CRB)~\cite{cover2012elements}. For any parameter vector $\theta$, the CRB
of $\theta$ is $\sigma_\theta^2 \geq \theta^T
g^{-1}\theta$, where the Information matrix (FIM)
\begin{equation}
    g_{\mu\nu} = \left\langle \frac{\partial^2 \log p}
        {\partial\theta_\mu\partial \theta_\nu} \right \rangle.
    \label{eq:fim}
\end{equation} 
The posterior $p=p(\vdelta, I|\phi,\psi)$ is given by Eqn.~\ref{eq:posterior}
and $\theta_\mu$ are the parameters, i.e.  $\vdelta$ and $I$. We can calculate
the na\"ive CRB for image registration, assuming we know the underlying image
$I$, and that $\partial I/\partial x$ and $\partial I/\partial y$ are
uncorrelated, the smallest possible variance on the estimation of the
$x$-direction shift $\Delta_x$ is
\begin{equation}
    \sigma^2_{\Delta_x} \geq \sigma^2 \bigg/ \int \rmd^2\vx 
            \left(\frac{\partial I}{\partial x}\right)^2.
    \label{eq:1dcrb}
\end{equation}
In other words, if the data are very noisy or if the underlying image has no
features, it will be difficult to measure the shifts. Note that the CRB predicts
that the shift error will scale linearly with noise ($\sigma_{\Delta} \propto
\sigma$). We reiterate that this is the CRB of the shifts assuming knowledge of
the true image $I$. Since this is an unrealistic assumption for real data, we
call Eq.~\ref{eq:1dcrb} and its discrete analog the na\"ive CRB.  For previous
derivations and discussions of the na\"ive CRB for image registration,
see~\cite{robinson2004fundamental,yetik2006performance}.  When discussing the
CRB below we use the definition related to Eq.~\ref{eq:fim} and not the
intuitive result of Eq.~\ref{eq:1dcrb}.

\subsection{Deriving the standard method of image registration}

In an experiment we have no access to the latent
image $I$. We offer a new derivation of the standard method for overcoming this
by marginalizing, or integrating out $I$: 
\begin{equation}
    p(\vdelta|\phi, \psi)
    \propto \int \rmd I ~ p(\phi, \psi|\vdelta, I)p(I).
\end{equation}
If we assume that $p(I)\propto 1$, i.e. all images are equally likely, we can perform the
integral by first recognizing that $||\psi-T_\vdelta
I||^2=||T_{-\vdelta}\psi-I||^2$ if $T_\vdelta$ is a unitary transformation
(preserves the L2 norm). Transforming discrete data will require interpolation.
Linear, quadratic, cubic, bi-cubic, and other local interpolation schemes
previously studied for this
problem~\cite{rohde2009interpolation,schreier2000systematic,inglada2007analysis,bailey2005bias}
are not unitary---neatly explaining some of their observed bias. In this work we
will consider only unitary interpolation by using Fourier shifting, however our
ultimate solution will obviate this discussion by directly employing
Eq.~\ref{eq:likelihood}. Now the posterior $p(\vdelta|\phi,\psi)$ is a product
of integrals of the form 
\begin{equation}
    \int \rmd x e^{-\frac{1}{2\sigma^2}\left((x-a)^2+(x-b)^2\right)} 
        \propto \exp\left(-\frac{(a-b)^2}{4\sigma^2}\right).
    \label{eq:marg-eqn}
\end{equation}
Applying this to each pixel in the data we arrive at the marginal likelihood 
\begin{equation}
    p(\vdelta|\phi, \psi) 
    \propto \exp \left(-\frac{1}{4\sigma^2} ||\psi - T_{-\vdelta}\phi||^2\right). 
    \label{eq:marg_likelihood}
\end{equation}

We have derived the standard least-squares similarity measure (it is
usually written down intuitively), in which one simply shifts one image until it
most closely matches the other. This process is illustrated by
Fig.\ref{fig:schematics}(a), which shows a pair of synthetic data which will serve as
$I$ in our numerical studies of periodic registration. It was calculated by
sampling a 64$\times$64 image from a power law in Fourier space 
\begin{equation}
    P(|I(k)|)\sim k^{-1.8} e^{-\frac{1}{2}\left(\frac{k}{k_c}\right)^2},
    \label{eq:idist}
\end{equation}
damped by a Gaussian with scale $k_c=k_\mathrm{Nyquist}/3$ to ensure a smooth
cutoff approaching the Nyquist limit, preventing aliasing.

Notice that if $T_\Delta$ is not unitary that this
objective is different depending on whether you shift one measured image or the
other. Note also that in general image registration this inverse transformation
may not exist; in such cases this method will fail. The literature features
multiple implementations of Eq.~\ref{eq:marg_likelihood} using Fourier
interpolation by either shifting the data~\cite{jacovitti1993discrete} or
upsampling by padding in Fourier space and finding the maximum
cross-correlation~\cite{guizar2008efficient}. The latter method can only be
as accurate as the factor of upsampling, e.g. quadrupling (in 2$D$) the number
of Fourier modes allows evaluating shifts of half a pixel. While sophisticated
extrapolations have been used to overcome the arbitrary choice of how much to
upscale, we will exactly shift the data and optimize
Eqn.~\ref{eq:marg_likelihood_estimator} directly. Writing the 2$D$ Fourier
transform operator as $\mathcal{F}$, we implement $T_\vdelta\phi$ as:
\begin{equation}
    T_\vdelta\phi = \mathcal{F}^{-1} e^{-i\mathbf{k}\cdot\vdelta} \mathcal{F}\phi
    \label{eq:shift}
\end{equation}

Another important result of our theory is the $4\sigma^2=(2\sigma)^2$ in the
denominator of Eq.~\ref{eq:marg_likelihood}: this likelihood function is for
data with twice the variance of our original problem, which is consistent with
taking the difference of two noisy signals. Some of the reported discrepancy
($\sqrt{2} \sim 40\%$) between the CRB and observed
error~\cite{robinson2004fundamental,aguerrebere2016fundamental,yetik2006performance}
can be explained by the absence of this factor. Those studying multi-image
registration have also neglected this modification of the noise fluctuations in
their estimating of shift precision~\cite{aguerrebere2016fundamental}.  
We have obtained by integrating out the latent image $I$ a distribution which
depends only on our data $\phi$ and $\psi$ and the unknown shift $\vdelta$.  We
can now define $\vdelta^\star_m$, the marginal maximum likelihood (ML) solution,
which we will now refer to as the standard Fourier shift (FS) method:
\begin{equation}
    \vdelta^\star_m = \text{max}_\vdelta p(\vdelta | \phi, \psi) 
                    = \text{min}_\vdelta || \psi - T_{-\vdelta} \phi ||^2.
    \label{eq:marg_likelihood_estimator}
\end{equation}

This new derivation of the standard method of image registration highlights and
clarifies some important limitations. Only unitary (L2-preserving) interpolation
for shifting images will lead to unbiased shift estimation, otherwise we are
simply optimizing a corrupted likelihood. Second, comparing the squared error
between shifted images is only correct if the noise in the images is Gaussian.
If we were studying images with Poisson-distributed noise, for instance, the
likelihood in Eqn.~\ref{eq:likelihood} should be a Poisson distribution. The
standard method is often successfully employed for non-Gaussian noise. We do not
doubt its efficacy, but instead claim that the standard method cannot be optimal
in this case because it violates the implicit assumptions of Gaussian noise.

\newcommand{\tpsi}{\widetilde{\psi}}
\newcommand{\tphi}{\widetilde{\phi}}
\newcommand{\ml}{\mathcal{L}}

\section{Statistical properties of the standard method}

\begin{figure}
    \includegraphics[width=0.95\columnwidth]{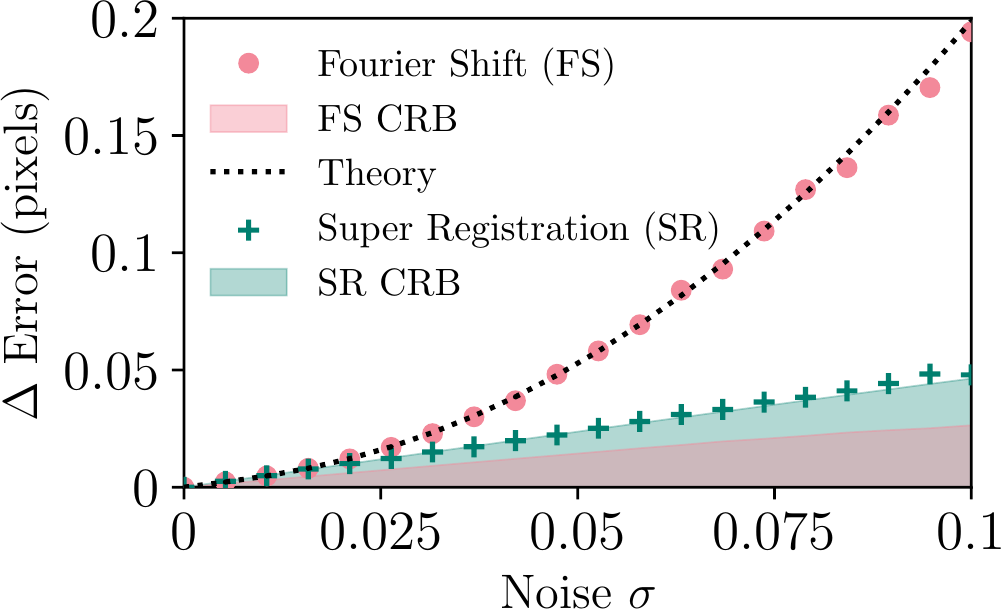}

    \caption{Comparing the noise-averaged errors of the inferred shift $\Delta$
    measured by the standard Fourier Shift method and Super Registration in the
    case of aligning synthetic periodic images. For each noise level, we
    generate an ensemble of 1000 $64\times 64$ images statistically similar to
    Fig.~\ref{fig:schematics} ($I(k) \sim k^{-1.8}$), measuring the average error 
	for both methods, along with the minimum expected error, CRB. 
	The error of the standard method (pink dots) grows
    quadratically with noise, whereas the naive CRB (pink shaded region)
    predicts a linear relationship.  Our theory (black dashed line) accurately
    describes the quadratic dependence in the error, matching numerical
    experiments. Super Registration (green pluses) demonstrates much lower
    error, recovers the linear relationship between error and noise, and reaches
    its CRB (green shaded region).}

    \label{fig:periodic_error}
\end{figure}

It is well documented in the literature that the errors in shift inference via
FS are much larger than the na\"ive CRB. Figure~\ref{fig:periodic_error} shows
the noise-averaged error (pink dots) of inferring the shifts as measured using
the standard Fourier shift method in Eq.~\ref{eq:marg_likelihood_estimator}.
The measured error grows quadratically with the Gaussian additive noise
$\sigma$, dwarfing The na\"ive CRB (shaded pink region). The follow section will
derive a theory (black dotted) to predict this quadratic error growth.

Say we measure the fields $\psi_i$ and $\phi_i$, then the log-marginal posterior
is (up to a constant) proportional to
\begin{equation}
    \ml = \frac{1}{2}\sum_i (\psi_i - T_{-\Delta} \phi_i)^2
        = \frac{1}{2} \sum_k |\tpsi_k - e^{ik\Delta}\tphi_k|^2,
    \label{eq:discrete_marg_likelihood}
\end{equation}
where $\tphi_k$ and $\tpsi_k$ are the Fourier transforms of our data. Our
measurements fluctuate around the true latent image $I$ according to 
\begin{align}
    p(\psi) &\propto \exp\left( -\frac{1}{2\sigma^2} 
        || \psi - I(x)||^2\right),\nonumber\\
    p(\phi) &\propto \exp\left( -\frac{1}{2\sigma^2} 
        || \phi - I(x-\Delta_0)||^2\right),
    \label{eq:datadist}
\end{align}
where $\Delta_0$ is the latent shift and $\sigma^2$ is the variance of the noise.
Near the true shift $\Delta_0$ we can expand the marginal likelihood as
\begin{equation}
    \ml(\Delta) = \ml(\Delta_0) + (\Delta-\Delta_0)\frac{\partial
    \ml}{\partial\Delta} + \frac{1}{2}(\Delta-\Delta_0)^2\frac{\partial^2
    \ml}{\partial\Delta^2} + \ldots,
\end{equation}
which is approximately minimized by
\begin{equation}
    \Delta-\Delta_0 = -\frac{\partial \ml}{\partial\Delta}\bigg/
                       \frac{\partial^2 \ml}{\partial\Delta^2}
                    = -i \frac{\sum_k k~\tpsi_k e^{-ik\Delta_0}\tphi_{-k}}
                    {\sum_k k^2~\tpsi_k e^{-ik\Delta_0}\tphi_{-k}}.
    \label{eq:generalerror}
\end{equation}
We can calculate the error 
of the standard method by averaging Eqn.~\ref{eq:generalerror} and its square
over the distributions in Eqn.~\ref{eq:datadist}.

\subsection{Bias of the standard method (1D)}

Writing Eqn.~\ref{eq:generalerror} as $A/B$ we can Taylor
expand about $A=\langle A\rangle$ and $B=\langle B\rangle$,
then average over the noise to find
\begin{equation}
	\left\langle \frac{A}{B}\right\rangle =
		\frac{\langle A\rangle}{\langle B\rangle}\left(1+
		  \frac{\mathrm{var}(B)}{\langle B\rangle^2}\right)
		- \frac{\mathrm{cov}(A,B)}{\langle B\rangle^2} + \ldots,
	\label{eq:ratioestimator}
\end{equation}
where $\langle\cdot\rangle$ denotes integration over the distributions of
Eqn.~\ref{eq:datadist}. Notice that
\begin{equation}
	\langle A\rangle = \left\langle \sum_k k 
		\tphi_k e^{-i k \Delta_0}\tpsi_{-k}\right\rangle = 
        \cancelto{0}{\sum_k k I_k I_{-k}},
    \label{eq:averagex}
\end{equation}
which is zero because the summand is odd in $k$. Therefore the average bias
for periodic images is to lowest order
\begin{equation}
    \left\langle \frac{A}{B}\right\rangle = 
	    -\frac{\langle AB\rangle}{\langle B\rangle^2}.
	\label{eq:ratioaverage}
\end{equation}
In general for non-periodic images $\langle A\rangle \neq 0$. Examining the
continuum limit of $\langle A\rangle$ in real space, we find
\begin{align}
    \langle A\rangle &= 
        \int \rmd x~ I \frac{\partial I}{\partial x}\nonumber\\
    &= \frac{1}{2}\int \rmd x \frac{\partial}{\partial x} I^2 
        = \frac{1}{2}\left(I(x_N)^2 - I(x_0)^2\right),
\end{align}
where $x_N$ and $x_0$ are the endpoints of the domain; $\langle A\rangle$ is a
total derivative depending only on the edges of the image. Therefore we hypothesize that
the bias of the standard FS method of image registration shown in
Fig.~\ref{fig:generalbiaserror} will be dominated by the edges of the data. Ziv
and Zakai in 1969~\cite{ziv1969some} and
others~\cite{robinson2004fundamental,nicholson1998bayesian}, share this
speculation, however, whereas they
argued that impingement of shift fluctuations onto the limits of the domain
caused bias, our theory suggests that structures of the edges of images
themselves cause bias.

Evaluating the remaining moments of Eq.~\ref{eq:ratioaverage} we find
\begin{equation}
	\langle B\rangle = \sum_k k^2 I_kI_{-k},
	\label{eq:bmean}
\end{equation}
which is the roughness of the latent image $I$, found in the denominator of the na\"ive CRB in
Eqn.~\ref{eq:1dcrb}. The last correlation for the average bias is
\newcommand{\kpr}{{k'}}
\begin{equation}
    \langle AB\rangle = \sum_{k\kpr} k {k'}^2~
        e^{-i(k+k')\Delta_0}\langle\tpsi_k \tpsi_{k'}\rangle
        \langle\tphi_{-k} \tphi_{-k'} \rangle,
	\label{eq:corrxystart}
\end{equation}
which can be evaluated using the moments
\begin{align}
    \langle \tpsi_k\rangle &= I_k, &\langle \tphi_k\rangle &= e^{-ik\Delta_0}I_k,\\
    \langle \tpsi_k\tpsi_k\rangle &= I_k I_k,
    &\langle \tphi_k\tphi_k\rangle &= e^{-ik2\Delta_0}I_k I_k,\\
    \langle \tpsi_k\tpsi_{-k}\rangle &= I_k I_{-k} + \sigma^2,
    &\langle \tphi_k\tphi_{-k}\rangle &= I_k I_{-k} + \sigma^2.
\end{align}
Considering the sum in Eqn.~\ref{eq:corrxystart} in three
cases $k'=-k$, $k'=k$ and $k'\neq \pm k$ we can apply the moments to find
\begin{align}
    \langle AB\rangle = \sum_k \Big(&k^3 \left((I_kI_{-k} + \sigma^2)^2 + 
                                (I_kI_{-k})^2 \right) +\nonumber\\
            &k \sum_{k'\neq \pm k} {k'}^2 (I_kI_{-k})^2\Big) = 0,
\end{align}
from which we conclude the entire correlation function vanishes due to each term of
the summand being odd in $k$. Further, numerical evidence and inspection of
higher order terms in the expansion of Eq.~\ref{eq:ratioestimator} support the
conclusion that for periodic images the standard Fourier shift method of image
registration is unbiased.

\subsection{Variance of the standard method (1D)}

Turning our attention to the variance or expected error of the bias given by
Eq.~\ref{eq:generalerror}; an expansion and average of $(A/B)^2$ (simplifying
for $\langle A\rangle = 0$) yields to lowest order
\begin{equation}
    \mathrm{var}\left(\frac{A}{B}\right) = 
        \frac{\langle A^2\rangle}{\langle B\rangle^2}.
    \label{eq:varab}
\end{equation}
Equation~\ref{eq:bmean} gives us $\langle B\rangle$, so we need only to compute the
correlation function $\langle A^2\rangle$:
\begin{align}
    \langle A^2\rangle &= -\sum_k \sum_{k'} k k' e^{-i(k+k')\Delta_0}
                    \langle \tpsi_k \tpsi_{k'} \rangle
                    \langle \tphi_{-k} \tphi_{-k'} \rangle\nonumber\\
    &= -\cancelto{0}{\sum_k\sum_{k'\neq k} kk' |I_k|^2|I_{k'}|^2}\nonumber\\
    &~~~~ +\sum_k k^2 \left((I_kI_{-k}+\sigma^2)^2 - (I_kI_{-k})^2\right),
    \label{eq:asquared}
\end{align}
where as before we have decomposed the sum into terms for which $k'\neq k$,
$k'=-k$ and $k'=k$. We find that the variance of the bias (which is also the
variance of the estimated shifts since we have shown $\langle \Delta \rangle =
\Delta_0$) is approximately
\begin{equation}
    \sigma^2_\Delta = 
    \left\langle(\Delta-\Delta_0)^2\right\rangle = 
        2 \frac{\sigma^2}{D^2}
        + \frac{L\pi^2}{3}\frac{\sigma^4}{D^4},
    \label{eq:biasvariance}
\end{equation}
where $D^2=\sum_k k^2 I_kI_{-k}$ is the roughness of the image. We used the fact
that $\sum_k k^2 = (2+L^2)\pi^2/3L\approx L\pi^2/3$ for a one-dimensional signal
with $L$ points. The lowest order term in Eq.~\ref{eq:biasvariance} is twice
the na\"ive CRB shown in Eq.~\ref{eq:1dcrb}, consistent with the fact that the
marginal posterior in Eq.~\ref{eq:marg_likelihood} has twice the variance of
the noise. We have shown that the standard Fourier shift method cannot achieve
the na\"ive CRB. Notice that the variance grows beyond the CRB at a rate
proportional to $\sigma^4$ and the image size $L$, so that error grows
quadratically with noise. This extra factor of the image volume means
that sampling a band-limited (sampled below the Nyquist limit) image at a higher
rate---increasing the resolution without increasing information content---can
actually decrease the registration precision for the standard Fourier shift
method. We discuss and verify this observation following an extension of this
theory to two-dimensions.

\subsection{Variance of the standard method in two dimensions}

Generalizing our expansion of the marginal likelihood we find
\begin{align}
    \ml(\vdelta) = &\ml(\vdelta_0) + (\vdelta-\vdelta_0)^T \nabla\ml\nonumber\\
                   &+\frac{1}{2}(\vdelta-\vdelta_0)^T~ \nabla^2\ml ~(\vdelta-\vdelta_0)
                   + \ldots,
\end{align}
from which we conclude that the two-dimensional analogue of
Eq.~\ref{eq:generalerror} is
\begin{equation}
    \vdelta-\vdelta_0 = - \left(\nabla^2\ml\right)^{-1}\nabla\ml.
    \label{eq:twodbias}
\end{equation}
If the off-diagonal terms of the Hessian $\nabla^2\ml$ are small compared to the
diagonal terms (the image is approximately isotropic), the two dimensions
decouple into an application of Eq.~\ref{eq:biasvariance} for each dimension. This
is generally a good approximation except for contrived data. In this case we
find the precision of two-dimensional image registration is approximately
\newcommand{\bk}{\mathbf{k}}
\begin{equation}
     \left\langle(\vdelta-\vdelta_0)^2\right\rangle = 
     \begin{pmatrix}
        2\frac{\sigma^2}{D_x^2}
         + \frac{N\pi^2}{3}\frac{\sigma^4}{D_x^4}\\[.5em]
        2\frac{\sigma^2}{D_y^2}
        + \frac{N\pi^2}{3}\frac{\sigma^4}{D_y^4}
    \end{pmatrix},
    \label{eq:twodvariance}
\end{equation}
where $N$ is the number of pixels in the one of the measured images, and $D_x =
\sum_\bk k_x^2 I_\bk I_{-\bk}$ and $D_y = \sum_\bk k_y^2 I_\bk I_{-\bk}$ are the
horizontal and vertical image roughness.  Eq.~\ref{eq:twodvariance} is used in
Fig.~\ref{fig:periodic_error} (black dotted) where we see excellent agreement
with the numerically measured error (pink dots). The excellent agreement---in
spite of ignoring the cross terms---can be explained by expanding
Eq.~\ref{eq:twodbias} for small values of the off-diagonal terms: the lowest
order correction averages to zero.

Our analysis has shown that the error of shift estimates of the standard Fourier
shift method grow much faster than the CRB. Why do the errors scale
quadratically with noise? Mackay found that in general and especially for
ill-posed problems (like distinguishing noise from signal), integrating over
parameters can yield distributions with stretched and skewed peaks, biasing the
maximum and leading to large errors~\cite{mackay1996hyperparameters}. We
integrated over all possible images in order to derive the standard FS
registration method.  Did this choice sabotage our effort to achieve the
ultimate precision? For exponential functions (like a Gaussian or our
likelihoods above), there is a deep relationship between optimization and
integration through Laplace's method or the method of steepest
descent~\cite{de1970asymptotic}. By integrating over all possible images, we
essentially maximized $\log p(\phi,\psi|I,\vdelta)$ over $I$---estimating the
latent image---and used that estimate for predicting the shift. This estimate is,
however, unreliable as it makes no distinction between the signal and the noise.
The high frequency modes of the data, dominated by noise and ironically most
discriminating for shift localization, cause the fluctuation of our inferred
shifts to be much larger than the CRB. This is illuminated by the following
section which considers the process of coarse-graining or binning image data.

\subsection{Coarse Graining Data can Improve Precision}

\begin{figure}
    \includegraphics[width=\columnwidth]{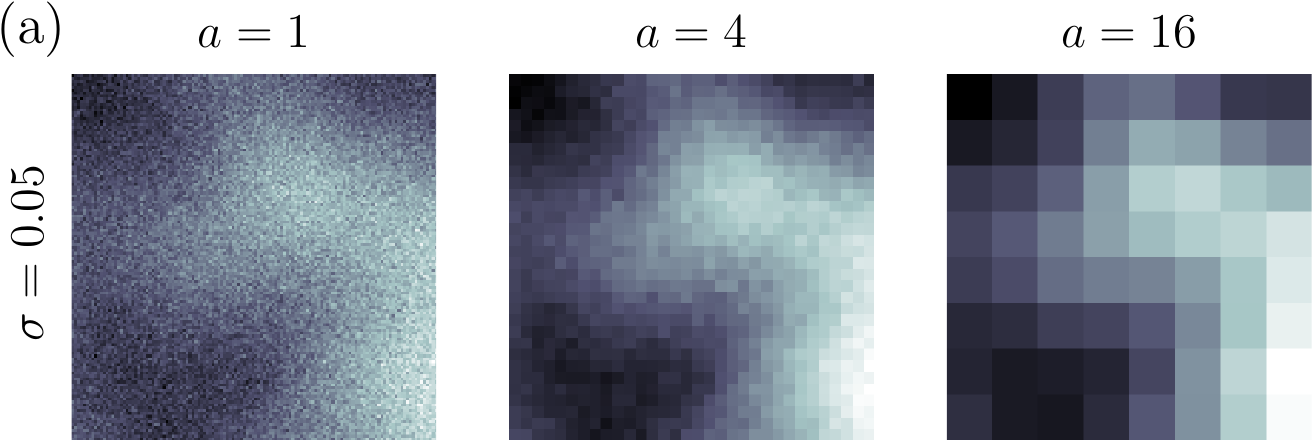}
    ~\\
    \includegraphics[width=\columnwidth]{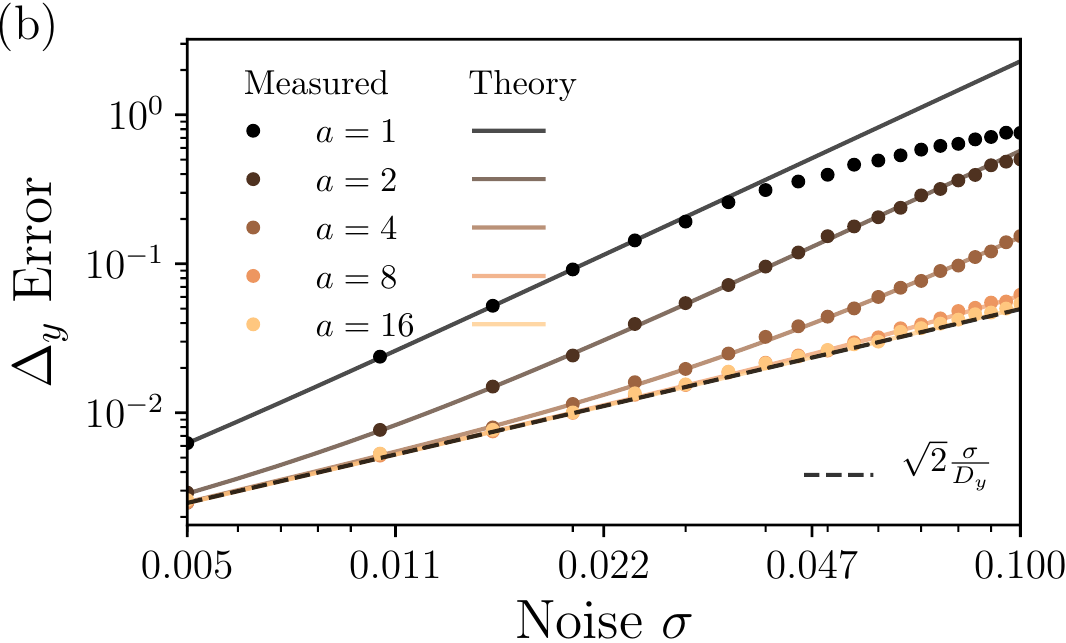}

    \caption{(a) An oversampled $1024^2$ image (the image varies on a scale $20\times$
    smaller than the Nyquist frequency limit) with 5\% additive white Gaussian
    noise then coarse-grained by summing over $a\times a$ blocks. Shown are
    $a=1$, $a=4$, and $a=16$, representing a drastic reduction in image size
    while not removing any information which localizes the shifts between
    images. (b) The error in inferred shifts (dots) for the standard Fourier
    shift method applied to the image after coarsening by 1, 2, 4, 8, and 16
    blocks. The original image was chosen to be smooth enough so that
    coarsening by a factor of 16 would not violate the Nyquist sampling theorem.
 	The solid lines are the prediction of our theory, and the dotted line is
    $\sqrt{2}$ times the na\"ive CRB, $\sqrt{2}\sigma/D_y$. }
    \label{fig:coarsen}
\end{figure}

Our theory for the variance of the shift predicts that $\sigma_\Delta^2 =
2\frac{\sigma^2}{D^2}\left(1+\frac{N\pi}{6}\frac{\sigma^2}{D^2}\right)$. The
factor of the image volume $N$ in the correction term inspired us to consider
reducing $N$ without changing $\sigma$ or $D^2$. Coarse-graining the data by
some linear factor $a$---shown schematically in
Fig~\ref{fig:coarsen}(a)---should not change the CRB assuming the latent image
$I$ is smooth on that length scale (or, equivalently, assuming that the data is
sampled at least $a$-times the Nyquist frequency).  Assuming that each pixel of
the data has noise of variance $\sigma^2$, the variance of noise for each
$a\times a$ block should be $a^2 \sigma^2$ (variances of uncorrelated noise
add). The denominator of the na\"ive CRB $D^2=\sum_k k^2 |I_k|^2$ is subtler:
the amplitude of each pixel increases by a factor of $a^2$ ($I_k\rightarrow a^2
I_k$), and the block sum only removed Fourier modes with zero amplitude by our
assumption above, so $D^2\rightarrow a^4D^2$.  Finally the coarse-grained image
will have its coordinates expanded by $a$, so that the variance should be
rescaled by $a^2$.  Therefore coarsening should modify our variance prediction
of the Fourier shift method accordingly:
\begin{align}
    \sigma^2_\Delta &= a^2\cdot 2 \frac{a^2 \sigma^2}{a^4D^2}\left(1+\frac{\pi
    N/a^2}{6}\frac{a^2\sigma^2}{a^4 D^2}\right)\nonumber\\
                    &= 2\frac{\sigma^2}{D^2}\left(1+\frac{\pi
    N}{6a^4}\frac{\sigma^2}{D^2}\right).
    \label{eq:coarsen}
\end{align}

Our theory predicts that coarse-graining over-sampled images can improve shift
inference by reducing the correction term, but that the method can at best yield
a variance equal to twice the naive CRB. This result may explain
improvements in registration precision from re-binning 
image intensities observed in other works~\cite{hutton2003software,pekin2017optimizing}.
Figure~\ref{fig:coarsen}(b) confirms the predicted 
relationship, where the black dots indicate the variance of a $N=1024^2$ image
which was oversampled by a factor of 20. Each lighter colored dot series
is the variance after coarsening by some factor $a$, and the solid lines are
given by Eq.~\ref{eq:coarsen}. We see excellent agreement with our theory, and
a convergence of the variances onto the $2\sigma^2/D^2$ line. Note that the
original image ($a=1$) variances differ from our theory for large noise: perhaps
the limits of large images and large noise are where our approximations in
truncating the Taylor expansion in Eq.~\ref{eq:varab} breaks down.

Coarsening smooth images only throws away information which is
dominated by noise. When we use the coarsened images in the standard FS method,
we implicitly estimate the underlying image but with less noisy modes, and will
get a more reliable estimate. In a real experiment without knowledge of the true
length scale of the image, we will not know the optimal coarsening length scale.
In the following section we propose our generative model which will use Bayesian model
selection to infer the image complexity supported by the data.

\section{Super registration}

How can we achieve the ultimate precision for image registration as predicted by
the CRB?  We have seen that the standard FS method of image registration which
directly compares two images has a variance in its shift prediction of the form
$\sigma^2_\Delta=2\sigma^2_\mathrm{CRB}(1+N\pi\sigma^2_\mathrm{CRB}/6)$, where
the CRB is $\sigma^2_\mathrm{CRB}=\sigma^2/\sum_k k^2 I_kI_{-k}$. We are still
studying periodic images, so it is natural to consider removing noise with a
filter like the optimal Wiener filter. This manifests by modifying our
log-marginal likelihood in Eq.\ref{eq:discrete_marg_likelihood} with the rule
$\tpsi_k\rightarrow A_k\tpsi_k$ and $\tphi_k\rightarrow A_k\tphi_k$, for some
filter function $A_k$.  This modification simply changes
$\sigma^2_\mathrm{CRB}\rightarrow\sigma^2/\sum_k k^2 A_kI_kA_{-k}I_{-k}$, and
since $A_kA_{-k}\leq 1$ (a filter only reduces power), this can only increase
$\sigma^2_\mathrm{CRB}$ and thus reduce our precision. 

Faced with this fact we abandon the standard method of image registration and
return to first principles by studying the likelihood defined by the image
formation model in Eq.~\ref{eq:likelihood}. Instead of shifting the data, we
will model the image and shift that, as shown schematically in
Fig.~\ref{fig:schematics}. This method will result in a de-noised and, depending
on the data, a super-resolution estimate of the latent image. Inspired by the
inextricable relationship between registration and super-resolution that we have
discovered, we call our new method Super Registration (SR). Our success depends
on using all that Bayesian inference has to offer, and so we proceed with a
discussion of evidence-based model selection.

\subsection{Bayesian inference and model selection}

Following Mackay's discussion on integration versus optimization in inference
with hyperparameters we will choose a model space and from this select the best
model by comparing the model evidence, $p(\phi,\psi)$. The evidence is simply
the normalization constant of the posterior Eq.~\ref{eq:posterior}; its
utility for selecting the best model can be exposed by a seemingly erudite
increase in notational complexity which makes manifest more of the assumptions
in our model. Consider a model of image formation for the case of periodic image
registration, expressed as the likelihood of measuring two images
$p(\phi,\psi|\vdelta,I)$. Now that we are optimizing over $I$ instead of
integrating, we must choose some parameterization $I\in\mathcal{H}$ where
$\mathcal{H}$ is some space of image models, e.g. a Fourier series or sums of
polynomials. This choice must be reflected in the conditionals of our
probabilities, so that the likelihood of measuring $\phi$ and $\psi$ must now be
written $p(\phi,\psi|\vdelta,I,\mathcal{H_\lambda})$, where
$\mathcal{H}_\lambda$ represents a specific choice of image model.

Proceeding with the inference task at hand by writing again (with our new
notation) the result of Bayes' theorem shown in Eq.~\ref{eq:posterior} we see
that the posterior now reads
\begin{equation}
    p(\vdelta,I|\phi,\psi,\mathcal{H}_\lambda) =
    \frac{p(\phi,\psi|\vdelta,I,\mathcal{H}_\lambda)p(\vdelta,I|\mathcal{H}_\lambda)}
        {p(\phi,\psi|\mathcal{H}_\lambda)}.
    \label{eq:modelposterior}
\end{equation}
The solution to our problem still lies in studying this posterior distribution,
but we now must also infer the best model $\mathcal{H}_\lambda$. We again apply
Bayes' theorem, finding the probability that our model is true given our
measured images
\begin{equation}
    p(\mathcal{H}_\lambda|\phi,\psi) \propto
    p(\phi,\psi|\mathcal{H}_\lambda)p(\mathcal{H}_\lambda).
    \label{eq:modelevidence}
\end{equation}
We have explicitly ignored the normalization constant $p(\phi,\psi)$~\footnote{
    \unexpanded{$p(\phi, \psi)=\sum_i
    p(\phi,\psi|\mathcal{H}_i)p(\mathcal{H}_i)$}.  This constant changes when we
    consider more models, which naturally must happen when we obtain more data,
    but does not influence the preference of one model over another.}.  
Assuming we have no prior preference for some models over others,
$p(\mathcal{H}_\lambda)\sim 1$, so inferring which model is most likely given
the data is equivalent to maximizing $p(\phi,\psi|\mathcal{H}_\lambda)$, which
is the normalization of Eq.~\ref{eq:modelposterior}.

Therefore Bayesian inference for image registration consists of the following
steps given some data $\phi$ and $\psi$.
\begin{enumerate}
    \item Choose some model $\mathcal{H}_\lambda$ and evaluate
        Eqn.~\ref{eq:modelposterior}, the
        posterior $p(\vdelta,I|\phi,\psi,\mathcal{H}_\lambda)$.
    \item Summarize the posterior by calculating the position and 
        widths of the maximum likelihood $\vdelta$ and $I$.
    \item Evaluate Eqn.~\ref{eq:modelevidence}, the model
        evidence $p(\phi,\psi|\mathcal{H}_\lambda)$, by estimating the
        normalization of the posterior.
    \item Repeat steps 1-3 with some subset of the model space $\mathcal{H}$.
    \item Choose the model $\mathcal{H}_\lambda$ with the largest evidence and
        examine its concomitant posterior distribution.
\end{enumerate}
The final (unlisted) step is to examine and decide whether the residuals and the maximum 
likelihood image and shifts are reasonable. 

This recursive process of acknowledging all the context and condition of our
model and inverting them with Bayes' theorem can go on forever. We could for
instance consider a probability over the parameters $\theta$ of our model
$\mathcal{H}_\lambda(\theta)$, adding another integration or optimization to the
steps above. Fortunately, the deeper these model assumptions
go, the less these decisions affect the outcome of our
inference~\cite{mackay1996hyperparameters}. Bayesian inference does not exclude
the experience of the researcher; we will terminate the inference recursion with
our own judgement. 

\subsection{Super Registration for periodic images}

Returning to our periodic image registration problem, let us pursue the
inference steps above in a concrete example. The natural model space for
periodic images consists of Fourier series, indexed by the maximum frequency
allowed. Given two images $\phi$ and $\psi$, the probability of measuring these
images given some latent image $I$ and shift $\vdelta$ is
\begin{align}
    \log p(\phi,\psi|\vdelta,I,\mathcal{H}_\lambda) = 
    -\frac{1}{2\sigma^2} \sum_{k=0}^\lambda &|\phi_k - I_k|^2 +\nonumber\\
    &|\psi_k - e^{-i\mathbf{k}\cdot\vdelta} I_k|^2\nonumber\\
    - \log Z_L &,
    \label{eq:periodicposterior}
\end{align}
where $\lambda$ indexes the complexity of the model and $\phi_k$, $\psi_k$, 
$I_k$ are the components of the Fourier transforms of our image model, and
$\mathcal{Z}_L$ is the normalization. Assuming a constant prior on shifts and
images, the maximum likelihood of the shifts and image is the solution of
\begin{equation}
    \vdelta_\mathrm{ML}, I_\mathrm{ML}=\mathrm{min}_{\vdelta,I}
        \sum_{k=0}^\lambda |\phi_k - I_k|^2 +
                    |\psi_k - e^{-i\mathbf{k}\cdot\vdelta} I_k|^2.
    \label{eq:mlperiodic}
\end{equation}

Equation~\ref{eq:mlperiodic} is in the standard form of a nonlinear least
square problem which we solve by alternating linear least squares for $I_k$ and
using Levenberg-Marquardt for $\vdelta$. For a given image model
$\mathcal{H}_\lambda$ we can find the most likely shift and image by evaluating
Eq.~\ref{eq:mlperiodic}, calculate the covariance, and compute the evidence.
Assuming flat priors on $\vdelta$, $I_k$ and $\mathcal{H}_\lambda$ the evidence
is the integral of our likelihood over $\vdelta$ and $I$:
\begin{equation}
    \mathcal{Z}_L = \int \rmd I_k \rmd \vdelta~
    p(\phi,\psi|\vdelta,I,\mathcal{H}_\lambda).
\end{equation}
$\mathcal{Z}_L$ can be computed by applying Laplace's method of integration using
the Jacobian of the least squares problem.

\begin{figure}
    \includegraphics[width=\columnwidth]{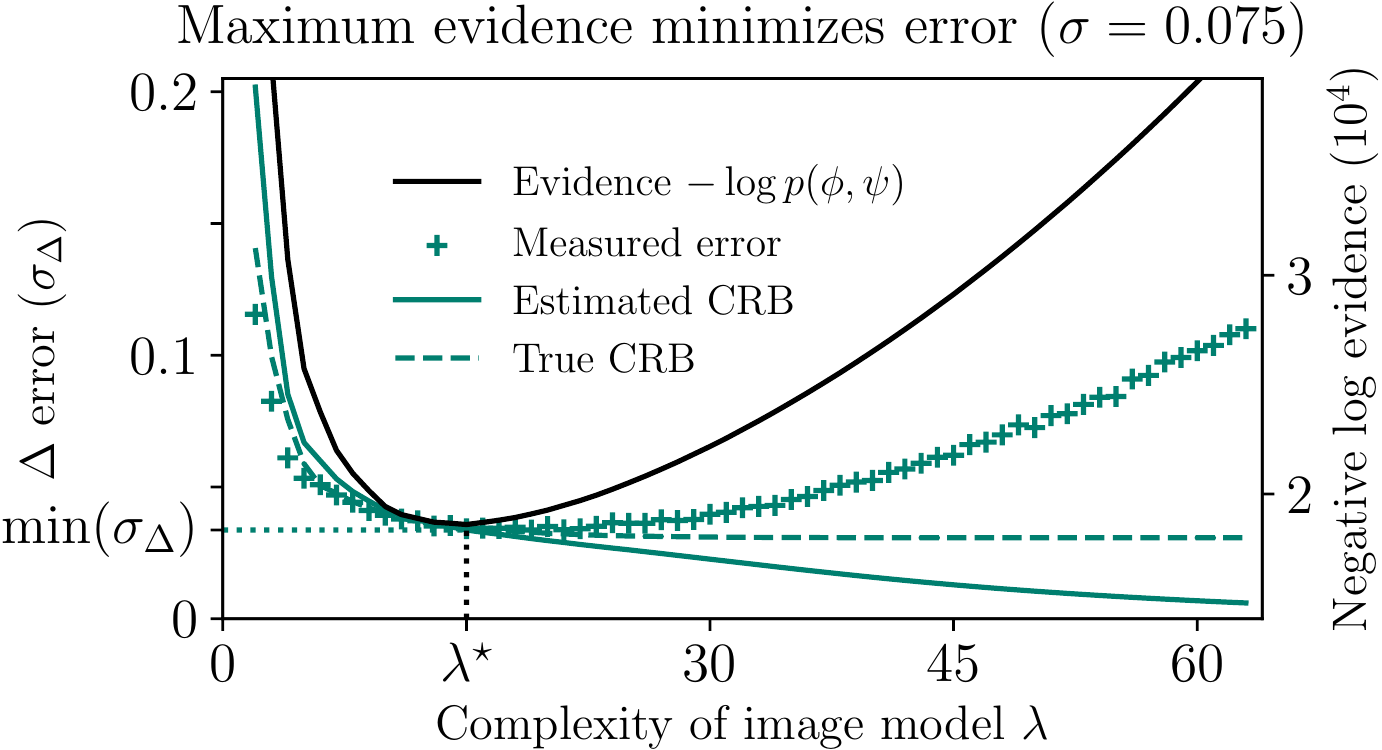}

    \caption{Using 1000 pairs of
    64$\times$64 images with additive Gaussian noise and $I(k)\sim k^{-1.8}$, we
    computed the model evidence $p(\mathcal{H}_\lambda|\phi,\psi)$
    (black curve) for all Fourier cutoffs indexed by $\lambda$, showing that
    when the evidence is maximized the actual shift error (green crosses) is
    minimized. Further, this error is nearly indistinguishable from the CRB
    (green dashed). Finally, the na\"ive estimate of the CRB (solid green) is
    computed from the curvature of the posterior using Eqn~\ref{eq:fim} the
    Fisher Information. During a real experiment only the
    evidence (black curve) and the na\"ive curvature estimate of the CRB (solid
    green) are available, but when the evidence is maximized all estimate of the
    error match.  }
    \label{fig:maxevidence}
\end{figure}

Figure~\ref{fig:maxevidence} shows the result of step 4 of our algorithm for the
periodic data used in all numerical experiments so far (shown in
Fig.~\ref{fig:schematics}), where have used every possible Fourier cutoff. We
have inverted the evidence to guide the eye, so that the minimum of the black
curve is the most likely model. For this true image and noise level the most
likely model is $\lambda=15$ (15$\times$15 sinusoids). The smallest observed
error (green crosses) in shift inference is also precisely at $\lambda=15$, and
is consistent with the CRB (green dashed). The most likely model provides the
most precise inference of the shifts. The maximum evidence solution has been
interpreted to embody Occam's Razor that the simplest explanation is most
likely~\cite{balasubramanian1997statistical}. Therefore evidence-based model
selection can systematically infer the number of degrees of freedom as supported
by the data, avoiding over-fitting and larger errors than the CRB.

The solid green line of Fig.~\ref{fig:maxevidence} is the CRB estimated by evaluating the second
derivative of the log-likelihood; notice that this erroneously continues to
decrease with increasing complexity. In a real experiment we only have access
to the evidence (solid black line) and this curvature estimate of the CRB (solid
green line). The maximum evidence model is also
where all of our estimates of the shift error, motivating
further the utility of the evidence-based choice of model complexity.
Finally note that when the complexity is chosen to be 64 (or all Fourier
modes are used) the measured error $\sigma_\Delta\approx 0.1$. In
Fig.~\ref{fig:schematics}, when the noise is $\sigma=0.075$, the same as in the
evidence experiment above, the observed error of the standard FS method is also
$\sigma_\Delta\approx 0.1$. Therefore we see numerical correspondence between
integration over the underlying image and optimization without selecting model
complexity by considering the evidence.

\begin{figure*}
    \begin{center}
    \includegraphics[width=0.75\textwidth]{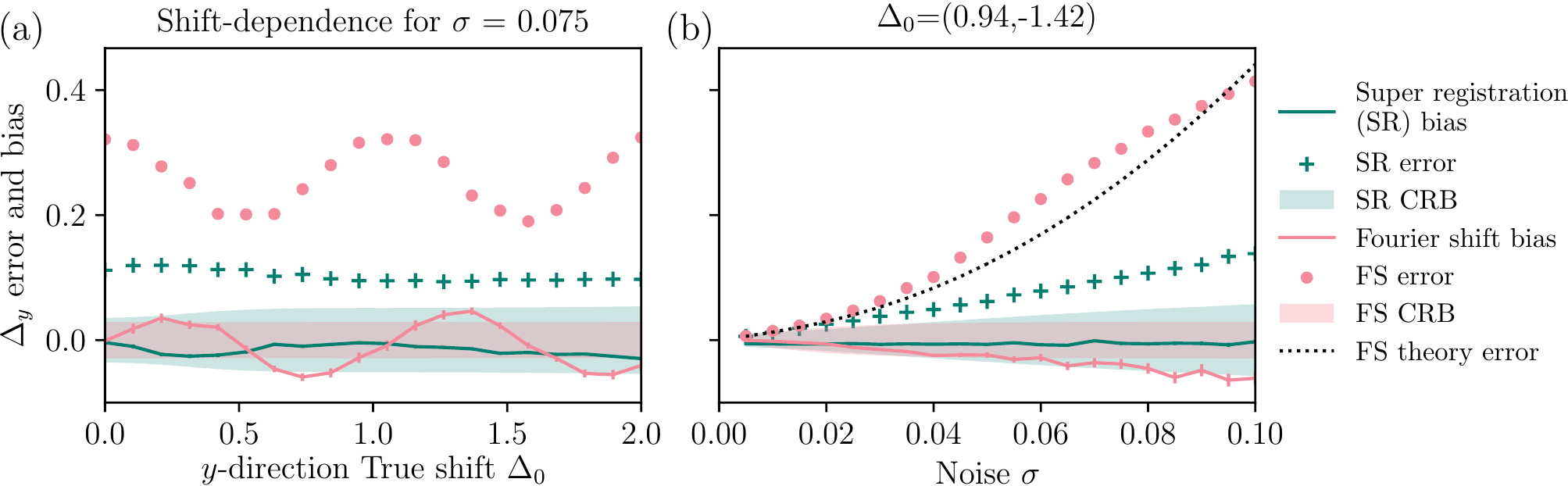}
    \end{center}

    \caption{Comparing the error and bias of the standard Fourier shift (FS)
    method and Super Registration (SR) for non-periodic data. The synthetic data
    were generated by the model $I(k)\sim k^{-1.8}$, twice as large as
    necessary, Fourier shifted and then cropped to produce non-periodic images.
    Errors and biases were measured from 1500 64$\times$64 noise samples.  (a)
    The $\Delta_y$ biases, errors, and CRBs for the standard FS (pink) and SR
    (green) are shown as a function of the true real shift $\Delta_0$. The
    standard method suffers from errors (pink dots) and bias (pink line) that
    are periodic in $\Delta_0$.  Super Registration shows almost zero bias
    (green line) and no periodic structure in the error (green crosses).
    Similarly to the periodic case, SR is much closer to its CRB (green shaded)
    than the standard FS method is to its CRB (pink shaded). (b) The biases,
    errors and CRBs for FS and SR methods as a function of noise for a fixed
    random shift $\Delta_0=(0.94, -1.42)$. The standard FS method has
    super-linear error (pink dots) growth with noise, and a monotonic bias (pink
    line) large than its CRB (pink shaded). Super Registration has linear error
    (green cross) growth about twice its CRB (green shaded), and a bias (green
    line) consistent with zero.}
    \label{fig:generalbiaserror}
\end{figure*}

\subsection{General non-periodic Super Registration}

Following the clarity of studying image registration in the periodic case, we turn
our attention to general non-periodic images. Here there is no clearly natural
model; images are extremely complicated. While there are exciting candidates in
the form of deep convolutional neural networks, these objects cannot (currently) be
evaluated at arbitrary points in space; they have no notion of continuous
locality~\cite{ulyanov2017deep}. In general the researcher's knowledge about
the physical objects being imaged should inspire the model space. A very
specific and successful example is the Parameter Extraction by Modeling Images
(PERI), which modeled almost every aspect of a confocal microscope, extracting
enough information from a light microscope to infer the parameters of the van
der Waals interaction~\cite{bierbaum2017light}.  Lacking such specific
inspiration therefore we chose sums of Chebyshev polynomials, in part because of
their excellent approximation properties~\cite{press2007numerical}.

We generated non-periodic data from the same distribution in
Eq.~\ref{eq:idist}, sampled twice as large (128$\times$128), shifted one by
$\vdelta_0$, cropped out a 64$\times$64 region, and added noise.
Figure~\ref{fig:generalbiaserror} show results for the error (pink dots and
green crosses) and bias (pink and green lines) using these synthetic data, as a
function of both noise $\sigma$ (Fig.~\ref{fig:generalbiaserror}(a)) and true
shift $\vdelta_0$ (Fig.~\ref{fig:generalbiaserror}(b)). Pink denotes the
standard FS method and green denotes Super Registration.
Figure~\ref{fig:generalbiaserror}(a) shows that the standard FS method has an
oscillating bias which is zero at whole and half-pixels, and has an oscillating
error which is largest at whole pixel shifts and smallest at half pixel shifts.
The pink shaded region is the CRB of the FS method.
Figure~\ref{fig:generalbiaserror}(b) shows super-linear error (pink) growth
for FS, compared with our theory from Eq.~\ref{eq:twodvariance} (black
dotted), and a bias (pink line) deviating slowly but consistently from zero.

Figure~\ref{fig:generalbiaserror}(a) shows that Super Registration has nearly a
constant bias (green line) and error (green crosses) as a function of true shift
$\vdelta_0$, and bias smaller its CRB (green shaded). The error is much smaller
than the standard FS method, and is one-third the error of the FS method when
$\sigma=0.1$ (10\% noise). Finally we see in Fig.~\ref{fig:generalbiaserror}(b)
that the error of SR grows linearly with noise. While SR here does not reach the
CRB, it scales the same as the CRB. A better image model should result in errors
more consistent with the CRB. Because we generated data by randomly sampling in
Fourier space, shifting, then cropping, our Chebyshev polynomials cannot
perfectly represent that signal.  This is an important reminder that the CRB
depends on the chosen model. Since the CRB is defined as the inverse of the
Fisher Information in Eq.~\ref{eq:fim}, the CRB is model-dependent, and thus the
standard FS method and SR have different bounds. 

How would Super Registration perform on data which has non-Gaussian noise?  We
cannot guarantee optimal precision in this case, because our model assumes the
noise is Gaussian. SR would provide reliable results, however, in the same way
that the FS standard method provides reliable results in this case. We can claim
this because optimization (SR) and integration (FS) are the same---following the
method of steepest descent or Laplace's method of integration---so that a fully
complex image model (one degree of freedom for each pixel) would be
statistically the same as shifting one image to match the other. The evidence
maximization procedure, however, is not guaranteed to be effective, as we know
the model assumes the incorrect noise distribution.

For many experimental images, Super Registration offers only a marginal
improvement in the image quality as measured by eye.  For a small shift error
$\vdelta-\vdelta_0$ the image intensity reconstruction error is $\Delta I
\approx (\vdelta-\vdelta_0)\cdot \vec{\nabla} I$. For smooth, highly sampled
images visual changes will be small. Most experiments do not operate in the
regime where they are not sampling at a high enough rate to see the structure of
their sample. Although the reconstructions for many experiments will not vary
dramatically visually, we show that the shift errors can dramatically
interfere with the information extracted from the reconstructions. When
inferring parameters from data such as object sizes, positions, and
orientations, correlation functions, and local contrast, the precision of these
quantities will be limited by the quality of the registration.  To emphasize the
scale of these errors, in the next section we demonstrate a dramatic improvement
in particle position inference from correctly registered images.

\section{Particle tracking errors}

\begin{figure}
    \includegraphics[width=\columnwidth]{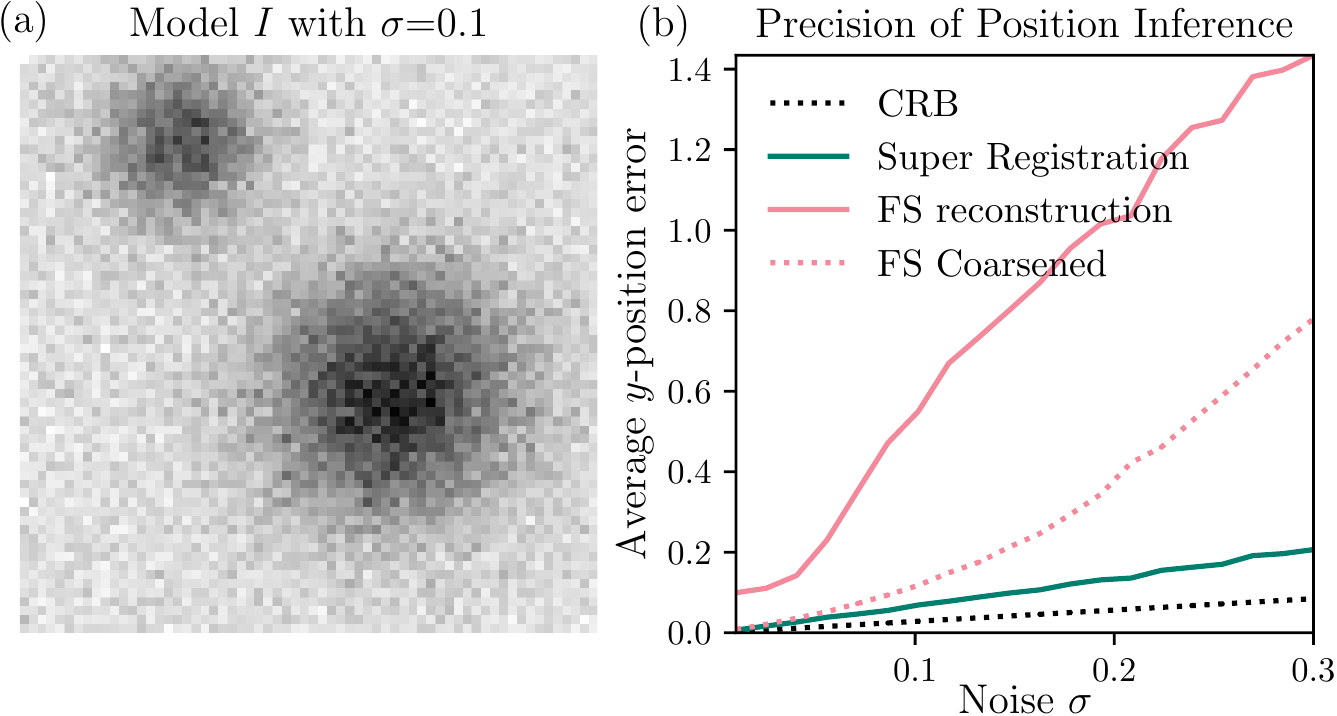}

    \caption{(a) A model image of two Gaussian particles with 10\% Gaussian
    additive noise. Eight of these images with random sub-pixel relative shifts
    were generated, and 1000 noise samples were drawn. For each noise sample,
    the underlying image was reconstructed either by the standard Fourier shift
    (FS) reconstruction or with Super Registration. With each 
    reconstruction we fit the Gaussian models which generated the data,
    inferring the most likely particle position and width.  (b) The average
    error of inferring the $y$-position of the larger particle from images
    reconstructed with the standard FS method (pink line), a coarse-grained
    image (pink dotted), and Super Registration (green line).}
    \label{fig:twop}
\end{figure}

A very common task in image processing is tracking particle positions. High
precision, especially in atomic-scale TEM and STEM, is important for
understanding real-space structure. For example, charge density waves cause
atoms to deviate from their lattice by tiny amounts, and can be studied by
carefully measuring the positions of the atoms in real
space~\cite{el2018nature}. For High-angle Annular Dark Field (HAADF) STEM, the
image of an atom is well-approximated by a 2D
Gaussian~\cite{yankovich2014picometre}. In TEM and STEM, noise is often
Poisson-distributed. Both SR and the standard method assume image noise is
Gaussian, and achieving optimalty for Poisson noise will require modeling the
noise correctly by modifying the likelihood in Eqn.~\ref{eq:likelihood}.
Assuming Gaussian noise, then, we created synthetic data of a pair of Gaussian
particles, shown in Fig.~\ref{fig:twop}(a) with 10\% additive noise.  Simulating
drift in a realistic STEM experiment, we created 8 copies of the two particle
images, randomly shifted. For each noise level we sampled 1000 noise instances,
with each reconstructing the underlying with both FS and our
Chebyshev-polynomial based Super Registration.

Figure.~\ref{fig:twop}(b) shows the error of inferring the position of the
larger particle using both the FS reconstruction (pink line) and SR
reconstructions (green line). For $\sigma=0.3$ or 30\% noise we see that the
precisions of particle position are 10x better using SR than FS. Further, the
SR method, not even using the correct model (a sum of Gaussian particles), is
only about twice the CRB for particle position inference (black dotted).
Finally, we show the result when using shifts inferred by the same data
coarsened by $a=3$, which was chosen to have the lowest error without being
biased. In summary we see that even though small shift errors do not have a
dramatic effect on the reconstructed image as measured by eye, there are
drastic effects on the precision of extractable information from the
reconstructions.

\subsection{Computational complexity}

The standard Fourier shift method requires a Fourier transform of one of the
images for each iteration of the optimization, ultimately scaling in time as
$\mathcal{O}(N\log N)$, where $N$ is the number of pixels in one image.  Super
registration requires estimating the underlying image, and thus requires $O(NM)$
where $M$ is the number of polynomials used in the image model. SR requires
trying multiple values of $M$, and multiple models, to find the greatest
evidence. For two $128\times 128$ images, FS takes less than a second on a
modern computer. SR requires an hour or more to try multiple values of $M$, but
only a few minutes to find the shifts and image for a given $M$. For
multi-image registration, optimal FS requires comparing all pairs of images,
and so scales as $\mathcal{O}(L^2N\log N)$ for $L$ images, whereas SR scales as
$\mathcal{O}(LNM)$, as it compares the data only to the model. Memory
requirements depend on the algorithm used. In this work we used
Levenberg-Marquardt nonlinear least-squares optimization, which requires
$\mathcal{O}(LNM)$ memory to store the Jacobian, and so images larger than
$128\times 128$ are impractical. 

There are several open opportunities for improving the performance of Super
Registration.  Memory consumption and computational time can be improved by
using Variational Inference and Stochastic Gradient Descent, which scales with
$\mathcal{O}(LN)$ in memory, and will be the subject of future work. A local
image model (where each image parameter only modifies a small area of the
image), such as radial basis functions, would scale even better than the Fast
Fourier Transform, as $\mathcal{O}(N)$. Finally, GPUs are designed to perform
optimal image calculations, and SR could achieve at least 10$\times$ (by na\"ive
FLOP counts) the performance as compared to a CPU.

\section{Conclusion}

Through a statistical theory of image formation, we have derived the standard
method of image registration, which shifts one image to match another. Our
theory predicts that shift errors for the standard FS method grow quadratically
with noise, much faster than the linear relationship of the CRB. Our explanation
for the deviation between the na\"ive CRB and the standard method comes from a
deep relationship between integration and optimization. The resulting formula is
useful for designing experiments which require image registration and must be
performed using the standard method. Our analysis leads to the surprising fact
that coarse-graining the data can improve the shift errors.

We develop a new method of image registration, which models the underlying
image, shifts that to match the data, and follows Bayesian inference to select
the image model for which there is the most evidence. Our theory reveals an
inextricable relationship between image registration and
super-resolution---that ultimate shift precision is predicated on selecting a
probable model. Therefore we named our new method Super Registration.  We
showed for periodic images that a Fourier series image model achieves errors
consistent with the CRB. We demonstrated superior bias and expected error
performance for general non-periodic images, and discussed the shortcomings of
our general model. Finally, we showed that, despite marginal improvements in image
quality as measured by eye, particle tracking experiments can be 10$\times$ more
precise when using Super Registration reconstructions.

Our results can be extended to more general transformations: by
application of the chain rule each term in our calculation of the average bias
and variance will be modified by partial derivatives. It is reasonable
to assume that the same problems---nonzero bias and errors which are much larger than
the CRB---will persist for transformations like affine skews, rotations, and
non-rigid registrations. Super Registration can accommodate all of these problems
by constructing the forward transformation instead of reconstructing the
inverse transformation.

Finally, medical imaging consists of lining up images of the same tissue from
different modes like X-ray and Magnetic Resonance Imaging
(MRI)~\cite{zollei2003unified,leventon1998multi}. The Super Registration method
involves constructing a generative model for the data, and this perspective
reminds us that contrast and features in X-ray and MRI will be different because
they respond to different underlying tissue structures.  Bias and large errors
for this problem have been observed and attributed to this
fact~\cite{tyler2018image}. Therefore some underlying model of tissue component
densities and a model of image formation (Super Registration) will be critical
for accurately and precisely registering these images.

Image registration is a very important and fundamental problem in medical
imaging, remote sensing, self-driving automobiles, non-destructive stress
measurement, microscopy, and more. Our theoretical study of the fundamental
problem of rigid shift registration in the presence of noise answers
long-standing questions on the precision and accuracy of shift inference, elucidates an
inextricable link between registration and super-resolution, and inspires a
solution to these problems with wide applicability.

\section*{Acknowledgements} Thanks to Ismail El Baggari, S.B. Kachuck, K.P.
O'Keeffe and D.B. Liarte for useful conversations in the preparation of this
manuscript. This work was supported by the NSF Center for Bright Beams, award
\#1549132.


\bibliographystyle{IEEEtran}
\bibliography{IEEEabrv,refs.bib}

\end{document}